\documentclass[prl,twocolumn,showpacs,showkeys,preprintnumbers]{revtex4}

\usepackage{amsmath}
\usepackage{amssymb}

\usepackage{graphicx}% Include figure files
\usepackage{epsfig}
\usepackage{rotating}

\usepackage{dcolumn}% Align table columns on decimal point
\usepackage{bm}% bold math
\begin{document}

%\preprint{\textsf{IFUP--TH 2005/??}}
%\preprint{\textsf{arXiv:nucl-th/05?????}}

\title{$^1S_0$ superfluid phase--transition in neutron matter with
realistic
nuclear potentials and modern many--body theories}

\author{Adelchi~Fabrocini}
\email{Adelchi.Fabrocini@df.unipi.it}
\affiliation{%
   Dipartimento di Fisica ``Enrico Fermi'', Universit\`a di Pisa
}%
\affiliation{%
   INFN, Sezione di Pisa, I-56100 Pisa, Italy
}%

\author{Stefano~Fantoni}
\email{fantoni@sissa.it}
\affiliation{%
   International School for Advanced Studies, SISSA
}%
\affiliation{%
   INFM DEMOCRITOS National Simulation Center,
   I-34014 Trieste, Italy
}%

\author{Alexei~Yu.~Illarionov}
\email{Alexei.Illarionov@pi.infn.it}
\affiliation{%
   INFN, Sezione di Pisa,
   I-56100 Pisa, Italy
}%
\affiliation{%
   Joint Institute for Nuclear Research,
   141980 Dubna, Moscow region, Russia
}%

\author{Kevin~E.~Schmidt}
\email{Kevin.Schmidt@asu.edu}
\affiliation{%
   Department of Physics and Astronomy,
   Arizona State University, Tempe, AZ, 85287
}%
\affiliation{%
   International School for Advanced Studies, SISSA,
   I-34014 Trieste, Italy
}%

\date{\today}% It is always \today, today,
             % but any date may be explicitly specified

\begin{abstract}
The $^1S_0$ pairing in neutron matter has been investigated in presence of
realistic two-- and three--nucleon interactions.
We have adopted the Argonne $v_{8^\prime}$ NN and the Urbana IX 3N
potentials.
Quantum Monte Carlo theory, specifically the Auxiliary Field Diffusion Monte
Carlo method, and Correlated Basis Function theory are employed in order to
get quantitative and reliable estimates of the gap.
They both fully take into account the medium modifications due to the
interaction induced correlations. The two methods are in good agreement
up to
the maximum gap density and both point to a slight reduction with respect
to the standard BCS value. In fact, the maximum gap is about
$2.5~\text{MeV}$ at $k_F \sim\, 0.8~\text{fm}^{-1}$ in BCS and
$2.3$--$2.4~\text{MeV}$ at $k_F \sim\, 0.6~\text{fm}^{-1}$ in correlated
matter. At higher densities the Quantum Monte Carlo gap becomes close to
BCS.
In general, the computed medium polarization effects are much smaller than
those previously estimated within \emph{all theories}.
Truncations of Argonne $v_{8^\prime}$ to simpler forms give the same
gaps in
BCS, provided the truncated potentials have been refitted to the same NN
data set. Differences among the models appear in the correlated theories,
most of the reduction being attributable to the tensor force.
The three--nucleon interaction provides an additional increase of the gap
of about 0.35 MeV.
\end{abstract}

%Use showpacs class option if PACS numbers display desired.

%Use showkeys class option if keyword display desired.

\maketitle

%\section{Introduction}

Superfluidity in highly asymmetrical nuclear matter, with a low
concentration
of protons, can lead to major consequences in modern astrophysics.
%
%The quantitative and realistic study of superfluidity in highly
%nuclear matter, with a low concentration of protons, has a capital
%relevance in modern astrophysics.
%
$^1S_0$ pairing may occur in the low density neutron gas in the inner crust
of neutron stars and in the proton phase in the interior of the star.
Anisotropic $^3P_2$--$^3F_2$ neutron pairing may also appear at higher
interior densities. Understanding microscopically such phenomena as the
cooling
rate~\cite{Tsuruta:1998,Heiselberg:2000} and the post--glitch relaxation
times~\cite{Sauls:1989,Pines:1992} requires as accurate as possible
knowledge
of the properties of the phases (normal and superfluid) of nuclear matter.

The role that tensor force and $N$--$N$ correlations have on the
superfluid
properties of neutron matter of $T = 0$ is not yet fully understood.
In this paper we address this problem by performing
\emph{ab initio} calculations of the  $^1S_0$ BCS--type pairing in 
neutron matter, using $(i)$ a realistic
Hamiltonian and $(ii)$ a trustworthy many--body theories. As far as the
Hamiltonian is concerned, a number of nucleon--nucleon interactions
have been derived in the last decade, all of them fitting the large
body ($\sim$ 3000) of the available $pp$ and $np$ scattering data with a
$\chi^2$ /datum $\sim 1$. The Argonne $v_{18}$~\cite{Argonne18},
Nijemegen~\cite{Nijemegen}, and CD Bonn~\cite{CDBonn} interactions 
are members of this
family of \emph{phase--equivalent} potentials. The situation is much less
established for the three--nucleon interaction (TNI) because of the much
smaller number of experimental data to be fitted as well as the larger
uncertainty of the theoretical framework on which its 
construction is based~\cite{VRP_TNI}..
%The question of which parts of the 3N potentials, as derived through
%standard
%quantum field theory, should be handled by inclusion of relativistic
%corrections, is currently still debated.
%
%$\chi^2$/datum $\sim 1$. To this family of $phase$ $equivalent$
%NN interactions belong the Argonne $v_{18}$~\cite{Argonne18},
%the Nijemegen~\cite{Nijemegen}, and the CD Bonn~\cite{CDBonn}
%potentials. Regarding the three--nucleon interaction (TNI), the situation
%is much less established, on one side because of the quite smaller number
%of experimental data to be fitted, and on the other for the larger
%uncertainty of the theoretical framework at the base of the construction
%of the 3N potential. A debated question of present days is whether
%and which part of the 3N potentials, as derived through standard quantum
%field theory, may be just a manifestation of relativistic effects.
%
%Within standard meson exchange potential theory, the present models of TNI
%contain the two--pion exchange contribution~\cite{Fuyita_TNI} as well as
%a number of extra terms coming from more complicated
%exchanges~\cite{VRP_TNI}.
%The parameters of the TNI are fitted to the light nuclei (A=3,4)
%experimental and nuclear matter empirical ground state properties.

There has been a rapid advance in dealing with strongly interacting systems
in their normal phase within modern many--body theories.
%The most accurate among these are the so called
%\emph{Quantum Monte Carlo} methods (QMC)~\cite{QMC_book}.
Parallel to the more traditional BHF and CBF theories there has been
significant development of the \emph{Quantum Monte Carlo} methods
(QMC)~\cite{QMC_book} in dealing with spin-dependent Hamiltonian.
In the different versions of QMC the normal phase many--body Schr\"odinger
equation is solved via stochastic sampling.
%In light nuclei~\cite{light_GFMC}, up to A=10, the solution is essentially
%exact, apart from statistical errors.
For homogeneous systems, a finite number of particle is put inside a
simulation box with periodic boundary conditions to simulate
the infinite gas. Then either the spatial configurations only
(Green's Function Monte Carlo, GFMC~\cite{QMC}) or both the
spatial and spin--isospin ones, by introducing auxiliary fields
(Auxiliary Field Diffusion Monte Carlo, AFDMC~\cite{AFDMC}),
are sampled. GFMC calculations are limited to a low number of
particles in the box (N=14 in neutron matter) since it sums over
all the 2$^N$ spin states of the N interacting neutrons, whereas AFDMC
does not have this limitation. As a consequence, GFMC has typically
large finite size corrections. GFMC~\cite{GFMC_nm} and AFDMC~\cite{AFDMC_nm}
have been so far used to study pure neutron matter with comparable results.
%It has to be stressed that both methods are not \emph{exact}, in principle,
%since they suffer of the fermionic \emph{sign problem}~\cite{QMC_book}.
Actual calculations restrict the sampling within the nodal surface of some
trial wave function, where the sampling guiding function has a constant sign
(Path Constraint approximation \cite{AFDMC_nm}).
An accurate choice of the nodal surface is essential in
obtaining good quality results~\cite{nodes_SF}. This is more difficult for
AFDMC since the method samples the spin degrees of freedom, and has less
inherent cancellation.

Correlated Basis Functions~\cite{feenberg} (CBF) theory provides an
alternative way of addressing interacting nuclear systems.
The strong nuclear interaction modifies the short range
structure of the wave function (short range correlations, SRC)
and introduces many--body contributions.
Within CBF the SRC are introduced through a many--body correlation
operator acting on a model function (Fermi gas for normal phase infinite
systems or shell model wave function for finite nuclei). Since the
NN potentials have important spin-- and isospin--dependent components,
the nuclear correlation operator is highly state dependent. An effective
correlation operator has been shown to be:
\begin{equation}
F_6(1,2,..N) \ = \ {\cal S} \left[ \prod_{i<j=1,N} f_6(ij) \right] \ ,
\label{eq:F6}
\end{equation}
where
\begin{equation}
f_6(ij) \ = \ \sum_{p=1,6} f^{(p)}(r_{ij}) O^{(p)}(ij) \ ,
\label{eq:f6}
\end{equation}
with $O^{(p=1,2,3)}(ij)=1$, ${\bf\sigma}(i)\cdot{\bf\sigma}(j)$,
$S(ij)=
(3 \hat r_\alpha(i) \hat r_\beta(j) - \delta_{\alpha \beta})
\sigma_\alpha(i)\sigma_\beta(j)$,
and
$O^{(p^\prime = p+3)}(ij) =
 O^{(p)}(ij) \times \left[\mathbf{\tau}(i) \cdot \mathbf{\tau}(j)\right]$,
$\mathcal{S}$ being a symmetrizer operator since the $f_6$ operators do not
commute among each other.
In pure $T = 1$ neutron matter, the isospin components may be actually
embodied into the spin dependent ones. This choice of the correlation
is consistent with the use of a momentum independent
$v_6$ potential:
\begin{equation}
v_6(ij) \ = \ \sum_{p=1,6} v^{(p)}(r_{ij}) O^{(p)}(ij) \ .
\label{eq:v6}
\end{equation}
Modern potentials also contain momentum dependent (spin--orbit and
so on) components. Accordingly, spin--orbit correlations have been used in
CBF nuclear matter studies~\cite{WFF}.
The correlation is variationally fixed by minimizing the ground
state energy as computed via cluster expansion.

The normal phase of neutron matter has been studied in details by
both QMC and CBF. Very recently
we have extended the two methods to deal with the $^1S_0$ superfluid
pairing. A detailed account of the theories is in preparation and
will be given in subsequent publications~\cite{QMC_BCS,CBF_BCS}.
We present here results obtained within the AFDMC and the CBF methods
for the gap energy computed starting from realistic potentials
(Argonne $v_{18}$ and its reductions plus the Urbana IX TNI)
and compare with the pure BCS results.
A brief outline of the methods will also be given.

 A correlated wave function for the neutron matter superfluid phase
is constructed as
\begin{equation}
|\Psi_s\rangle \ = \ \hat F \ |\text{BCS}\rangle \ ,
\label{eq:psi_s}
\end{equation}
where the model BCS--state vector is
\begin{equation}
|\text{BCS}\rangle \ = \ \prod_{\bm k}
( u_{\bm k} + v_{\bm k}a^\dagger_{{\bm k}\uparrow}
a^\dagger_{-{\bm k}\downarrow})|0\rangle \ .
\label{eq:BCS}
\end{equation}
$u_{\bm k}$ and $v_{\bm k}$ are real, variational BCS amplitudes,
satisfying the relation $u^2_{\bm k} + v^2_{\bm k} = 1$,
$|0\rangle$ is the vacuum state and $a^\dagger_m$ is the fermion
creation operator in the single--particle state, whose wave function is
\begin{equation}
\phi_{m\equiv {\bm k}, \sigma}(x\equiv {\bm r},s) \ = \
 \dfrac{1}{\sqrt{\Omega}} \
\eta_\sigma (s) \exp(\imath {\bm k}\cdot {\bm r}) \ .
\label{eq:single}
\end{equation}
$\Omega$ is the normalization volume and
$\eta_{\sigma=\uparrow,\downarrow}(s)$ is the spin wave function with
spin projection $\sigma$. The second--quantized correlation operator
$\hat F$ is written in terms of the N--particle correlation operators,
$\hat F_N$, as
\begin{equation}
\hat F \ = \ \sum_{N, m_N} \ \hat F_N \
|\Phi_N^{m_N}\rangle\langle\Phi_N^{m_N}| \ ,
\label{eq:hat_F}
\end{equation}
where $m_N$ specifies a set of single--particle states. In coordinate
representation and for a $f_6$--type correlation, we have:
\begin{gather}
\langle x_1,x_2,..x_N|\hat F_N | \Phi_N^{m_N}\rangle\ \ = \
\nonumber \\
F_6(1,2,..N)
\left\{ \phi_{m_1}(x_1)\phi_{m_2}(x_2)..\phi_{m_N}(x_N)\right\}_A \ .
\label{eq:F_N}
\end{gather}
The suffix $A$ stands for an antisymmetrized product of single--particle
wave functions.

The cluster expansion of the two--body distribution function, $g(r_{12})$,
and of the one--body density matrix, $n(r_{11^\prime})$, for a simple
Jastrow correlations ($f_J(i,j)=f^{(1)}(r_{ij})$) was developed in
Ref.~\cite{Fantoni:1981}. It is possible to sum, in a Fermi Hypernetted
Chain (FHNC) theory, all the cluster
diagrams contributing to the two quantities and constructed by the
dynamical correlation lines ($h_J = f_J^2 - 1$) and the by BCS
statistical correlations,
\begin{align}
l_v(r) =& \dfrac{\nu}{\rho_0} \int \dfrac{d^3k}{(2\pi)^3}
\ \exp ({\imath {\bm k} \cdot {\bm r}})\ v^2(k) \ ,
\label{def:l_v} \\
l_u(r) =& \dfrac{\nu}{\rho_0} \int \dfrac{d^3q}{(2\pi)^3}
\ \exp ({\imath {\bm k} \cdot {\bm r}})\ u(r) v(r) \ ,
\label{def:l_u}
\end{align}
where $\rho_0$ is the average density of the uncorrelated BCS model
and $\nu=2$ is the neutron matter spin degeneracy.
For more complicated correlations, like the $f_6$ model,
complete FHNC expansion cannot be derived. So, we have computed the
expectation values at low orders of their cluster expansions.
In particular, the average density,
\begin{equation}
\rho = \dfrac{\langle \hat{N}\rangle }{\Omega} \ =
\dfrac{\sum_m <a^{\dagger}_m a_m>}{\Omega} \ ,
\label{eq:OBD}
\end{equation}
is computed at the first order of the expansion in a series
of powers of the dynamical correlations.
 This expansion provides at each order the correct density normalization
in the normal phase, since it fully takes into account cancellations
between
same order diagrams but with different numbers and types of statistical
correlations. Similar, if not complete cancellations, hold in the
superfluid phase. Consistently, the matrix elements of the Hamiltonian
on the
correlated BCS state are evaluated at the two--body cluster level with
vertex corrections at the interacting pair.

In AFDMC for a BCS--like phase the neutrons are paired in a
pfaffian~\cite{QMC_BCS,Carlson:2003} constructed by a trial BCS amplitude.
To this aim, we have modified the normal phase AFDMC code, substituting
the Slater determinant with the pfaffian. We have taken the BCS amplitudes
from the CBF calculations. The AFDMC simulation has been performed for
N=12--18 neutrons in a periodic box and interacting through the
Argonne $v_{8^\prime}$~\cite{Argonne8prime} (A8') potential.
Preliminary results with larger
N--values confirm the findings of this paper.
Table~(\ref{tab1}) shows the results obtained by using a time step of
$\Delta\tau = 5\times10^{-5}~\text{MeV}^{-1}$.
The energy per particle for the normal phase at
$k_\text{F} = 0.6~\text{fm}^{-1}$ for the closed shell case, N=14,
is $E_{\text{NP}}/N = 2.548(3)~\text{MeV}$.
For even N--values all neutrons are paired in the pfaffian, whereas
for odd N--values the configuration of the unpaired neutron
providing the best energy must be found.
The gap energy for odd N--values is calculated according to:
\begin{equation}
 \Delta(N) = E(N) - \dfrac{1}{2}\left( E(N+1) + E(N-1) \right) \ .
\label{eq:delta}
\end{equation}

\begin{table}
\caption{%
 AFDMC energy per particle, $E/N$,
 and gap energy, $\Delta$, in MeV, in neutron
 matter with the $A8^\prime$ potential for 12$-$18 BCS--paired neutrons
 at $k_\text{F} = 0.6~\text{fm}^{-1}$.
% The normal phase, $E_\text{NP}$, energy is also given.
% $N_B$ is the number of blocks used in the simulations.
% Each blocks is made of 10 steps.
 In parentheses the statistical errors are given.
}%
\begin{ruledtabular}
\begin{tabular}{cccc}
 $N$ & $E/N$ & $E$ & $\Delta$ \\
\hline
 12 &2.6356(17) & 31.627(21) & \\
 13 &2.7593(17) & 35.871(22) & 2.182(37) \\
 14 &2.5536(15) & 35.750(21) & \\
 15 &2.8036(17) & 42.054(26) & 2.855(44) \\
 16 &2.6654(18) & 42.647(29) & \\
 17 &2.8075(15) & 47.727(25) & 2.333(49) \\
 18 &2.6746(17) & 48.142(31) & \\
\hline
 &&& \textbf{2.457(76)} \\
\end{tabular}
\end{ruledtabular}
\label{tab1}
\end{table}

The present calculations show that, at shell closure for N=14, the normal
phase is energetically slightly preferred to the superfluid one.
Two effects must be considered: improvements to the pairing functions
(e.g. a better choice of the nodal surface) and that the 14 particles
box dimensions are too small for the correlation length of the
pairing function. Again, we have indications that the superfluid
phase is actually preferred from preliminary calculations with larger
number of particles.

%\begin{figure}[ht]
%\includegraphics[angle=0, scale=0.5]{eps/Nk13.eps}
%\vspace{0.5cm}
%\caption[]{
%AFDMC energy for neutron matter at $k_\text{F} = 0.6~\text{fm}^{-1}$
%for 13 neutrons as a function of $k^2$ of the unpaired state.
%The potential used is $A8^{\prime}$.
%}
%\label{figk13}
%\end{figure}

\begin{figure}[ht]
\vspace{5.mm}
\includegraphics[angle=0, scale=0.38]{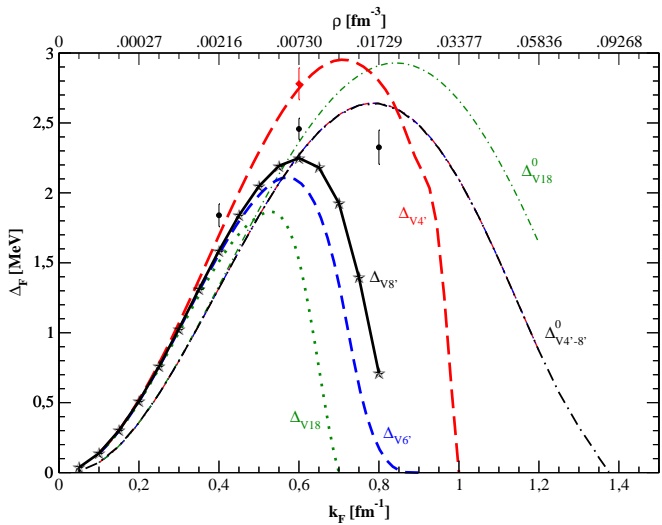}
\caption[]{\label{fig1}%
%(colored online)
$^1S_0$ gaps for different nucleon-nucleon potentials and methods.
 See text.
}%
\end{figure}

The results for the gap function at the Fermi momentum are collected
in Figure~(\ref{fig1}). The curves labeled $\Delta^0_{v_x}$
refer to the pure BCS estimates using operatorial reductions of 
the Argonne $v_{18}$ potential, down to the $v_{4}$ one. 
Most of the reduced potentials have been refitted to reproduce 
the $S$-- and $P$--wave
experimental phase shifts ($v_{4'}$--$v_{8'}$) . It is remarkable that all
these potentials give \emph{the same} BCS gap, provided they fit the
same data set. Presumably this is a consequence of getting the effective
 interaction near the Fermi surface correct.
 A version of Argonne $v_{18}$ simply cut to $v_{6}$, without refitting,
provides the higher BCS gaps of the $\Delta^0_{v_6}$ curve. In BCS, 
$\Delta^0_{v_6}=\Delta^0_{v_{18}}$.  
The $\Delta_{v_{4^\prime-8^\prime}}$ curves show the CBF gaps for the
corresponding refitted potentials.
The points with error bars are the AFDMC gap estimates
with Argonne $v_{8^\prime}$. 
The AFDMC results are the average values of the
gaps (\ref{eq:delta}) computed around N=13, 15 and 17.

The Figure shows that the low density (up to $k_\text{F} \sim
0.6~\text{fm}^{-1}$) CBF and AFDMC gaps, for the same A8' potential
model, are in good agreement. The low order cluster expansion provides
a CBF gap slightly smaller than the quantum MC one. The highest Fermi
momentum AFDMC result, at $k_\text{F} = 0.8~\text{fm}^{-1}$, is much higher
of the corresponding CBF gap, stressing the need of pushing the cluster
expansion to higher orders at increasing densities. It is clear, by
looking at the $\Delta_{v_{4^\prime}}$ and $\Delta_{v_{6^\prime-8^\prime}}$
curves, that most of the reduction of the gap with respect to BCS is due to
the tensor force, only partially compensated by the spin--orbit potential.
The CBF $\Delta_{v_6}$ is the lowest curve, pointing once more
to the importance of tensor force,
%to the importance of having a \emph{realistic} potential,
in the sense that
experimental data \emph{must} be fitted in order to obtain meaningful
results.

The reduction of the gap due to medium effects is much smaller
than previous estimates~\cite{Martino_RMP}, which provided suppressions
of a factor of two and more. The CBF theory shows an
early disappearance of the gap, around $\rho\sim 0.15$ $\rho_{NM}$,
$\rho_{NM}$=.16 fm$^{-3}$ being the empirical nuclear matter
saturation density. However, the QMC calculations do not seem to support
this finding, and, surprisingly enough, give a gap energy that is still
close to the standard BCS one.
A similar result was found in a preliminary study employing
GFMC and simple model potentials~\cite{GFMC_gap}.

%{\bf Comparison with Catania for Argonne 18.}

The square in Figure gives the AFDMC gap, at $k_\text{F} =
0.6~\text{fm}^{-1}$,
for the A8' model implemented by the Urbana IX three--nucleon
interaction~\cite{Argonne8prime}. The gap with the three--body force
results to be $\Delta_{v_{8^\prime}+UIX}$=2.810(146) MeV, slightly
increased
with respect to $\Delta_{v_{8^\prime}}$=2.457(76) MeV. A qualitatively
different
result was found in Ref.~\cite{Zuo:2002}, where the authors have obtained
a small decrease of the gap in a Bruckner $G$--matrix based approach.

In this letter we have used Quantum Monte Carlo and Correlated Basis
Functions theories to microscopically evaluate the $^1S_0$ superfluid
gap in pure neutron matter with modern interactions. These methods
allow for taking into account medium modification effects in a consistent
and realistic way. In particular, QMC is expected to give a solution
of the many--body Schr\"odinger equation very close to the true one
Both theories are in good agreement up to the maximum gap density,
and show a slight reduction of the maximum gap with respect to standard BCS.
At higher densities QMC gives a larger gap than CBF, probably because
the CBF gap is computed at low order of the cluster expansion and the
missing
diagrams become more and more relevant with the density. The gap
reduction is essentially due to the tensor interaction.
A novel and important effect is the small influence of the medium
polarization, contrary to all the previous estimates.
The three--nucleon interaction provides a small increase of the gap.
It will be most interesting to extend these analysis to other types
of pairing, as the $^3P_2-^3F_2$ neutron pairing and the
 $^1S_0$ proton pairing in highly asymmetrical nuclear matter.

%%%%%%%%%%%%%%%%%%%%%%%%%%%%%%%%%%%%%%%%%%%%%%%%%%%%%%%%%%%%%%%%%%%%%%%%%%%

%%%%%%%%%%%%%%%%%%%%%%%%%%%%%%%%%%%%%%%%%%%%%%%%%%%%%%%%%%%%%%%%%%%%%%%%%%%

\acknowledgments
A.Yu.I. acknowledges the support of INFN and of the Dipartimento di Fisica ``Enrico Fermi'' of the  Pisa University.

\bibliographystyle{revtex}
%\bibliography{bib/S-BCS}

\end{document}